# Cathode Materials for Lithium Ion Batteries (LIBs): A Review on Materialsrelated aspects towards High Energy Density LIBs

**Ambesh Dixit***

*Department of Physics & Center for Solar Energy, Indian Institute of Technology Jodhpur, Jodhpur 342037, India*

**ambesh@iitj.ac.in*

**Abstract**

This article reviews the development of cathode materials for secondary lithium ion batteries since its inception with the introduction of lithium cobalt oxide in early 1980s. The time has passed and numerous cathode materials are designed and developed to realize not only the enhanced capacity but also the power density simultaneously. However, there are numerous challenges such as the cyclic stability of cathode materials, their structural and thermal stability, higher operating voltage together with high ionic and electronic conductivity for efficient ion and charge transport during charging and discharging. This article will cover the development of materials in chronological order classifying as the lithium ion cathode materials in different generations. The ternary oxides such as LiTMOx (TM=Transition Metal) are considered as the first generation materials, whereas modified ternary and quaternary oxide systems are considered as the second generation materials. The current i.e. third generation includescomplex oxide systems with higher lithium content such as $Li_2TMSiO_4$ aiming for higher energy density. Further, developments are heading towards lithium metal based batteries with very high energy densities.

## Introduction:

The demand of energy consumption per person is continuously increasing not only because of comfort living but also to meet every day necessities such as mobiles, laptops, and other electronic gadgets for every day communication, commuting, power electronics, and house hold accessories especially in remote areas. Thus, batteries, particularly rechargeable ones, are becoming essential and integral part of everyone's life. Further, with these applications energy demand especially for movable platforms such as buses, rickshaws, cars, trains and even planes currently rely on uses of conventional non-renewable i.e. fossil fuels. The use of fossil fuels is also impacting our environment and severely polluting with various toxic and greenhouse gases, leading to adverse effect on living beings. That's' why there is a continuous compulsion to use renewable/green energy sources, such as solar, wind, thermal and geothermal, not only because of limited conventional fossil fuel sources depleting every day but also because of large pollution after consuming conventional fossil fuel products, causing environmental degradation and thus, life threatening consequences. Most of the megacities are facing such environmental degradation because of toxic exhaust gases from regular vehicles and other exhausts such as conventional power plants, burnings, and industries using conventional fuel sources. The situation is worse in developing and under developed countries, where there is no guidelines and policies on using conventional fossil fuels. The alarming situation is that the consumption of conventional fuel is increasing every day even after knowing the consequences. Further, there are intrinsic challenges with renewable energy sources such as their availability, e.g. Sun light is available during day time and energy requirement is equally or even more important for off-Sun hours, unpredictability, uncontrollability, and intermittency apart from day and night cycles. In spite of all these, the development of solar photovoltaics led to the installation of large scale PV power plants producing GW (Giga Watts) poweracross the globe and providing useful electrical energy from the freely available Sun light. This directly converted electrical energy can be either fed to the grid or used in distributed applications. However, after Sun hours, the generation of electricity is not possible and thus, there is a stringent requirement of storing electrical energy simultaneously for its use during off-Sun hours. This relies on the energy storage systems i.e. batteries in terms of large specific capacity and power density together with cyclability. The energy needs for various applications such asoff-Sun hour grid, distributed off grid, and small scale energy applications,can be met by using efficient electrical energy storage devices. Considering such constraints, large efforts are put in the development of secondary rechargeable batteries to meet some of these requirements.

The large energy storage requirements are currently met by lead acid batteries in general, which are not safe to handle because of lead as the toxic elements and also suffer from lower operating voltage and specific energy density, as illustrated schematically in Ragone plot, Fig 1.

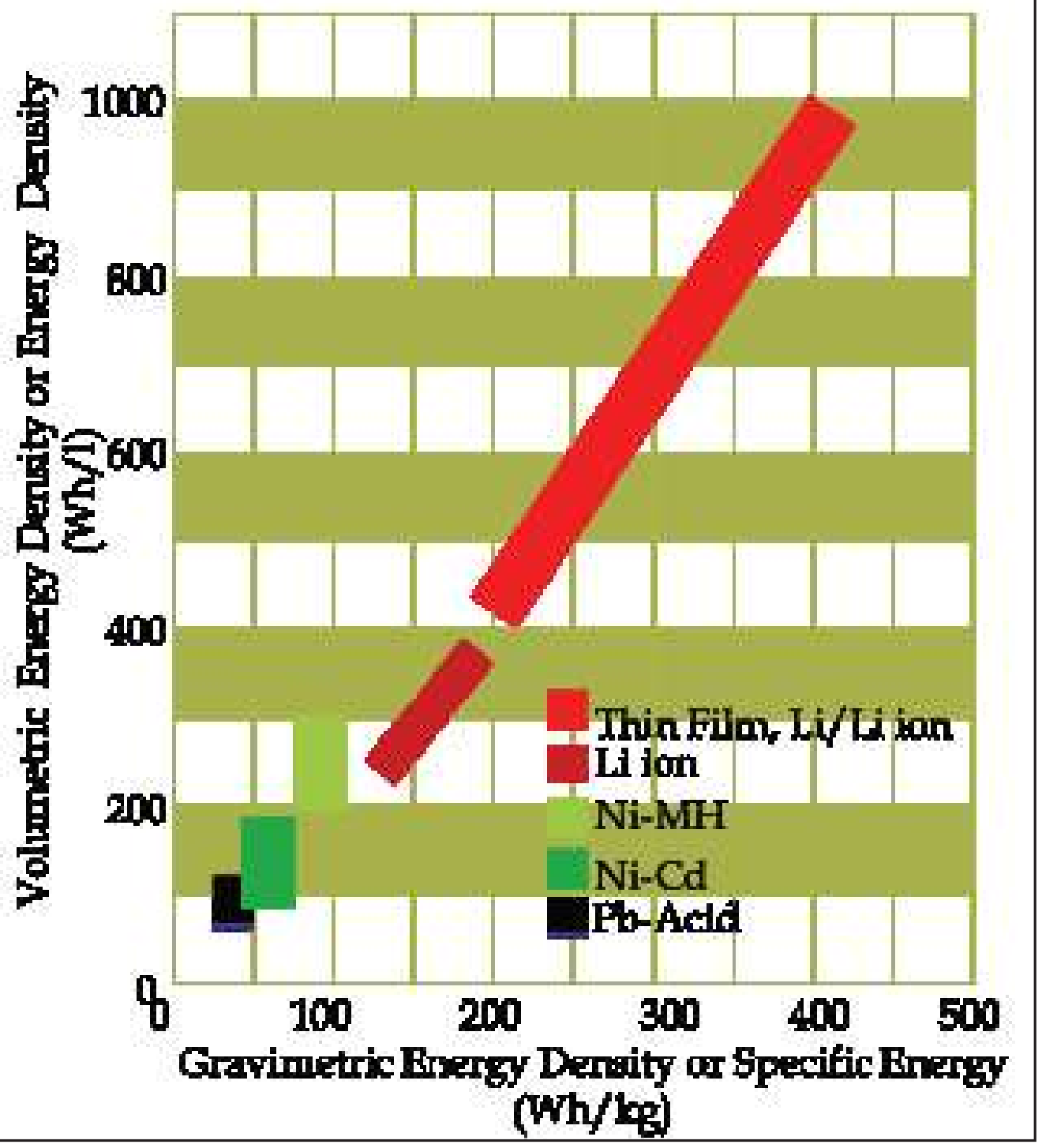

*Figure 1. Ragone plot, showing gravimetric and volumetric energy densities of different rechargeable batteries*

This provides opportunities to innovate materials which can offer high energy density, cyclability without capacity fading, together with safety against environmental factors. The lithium ion batteries are considered as one of the probable solution towards mitigating the intermittency of renewable energy sources by replacing the widely used toxic lead acid batteries. However, currently lithium ion rechargeable batteries are mostly used in portable electronic devices and in its matured state, which replaced the initially used nickel-cadmium rechargeable batteries. These Ni-Cd batteries were offering relatively poor energy and power density with respect to lithium ion batteries, Fig 1, and also suffering from memory issues, ending with relatively shorter life time. Thus, lithium ion batteries are far superior from their counter systems such as lead acid (Pb-acid), nickel-cadmium (Ni-Cd), and nickel-metal hydride (Ni-MH) batteries, Fig 1. The relatively high gravimetric and volumetric energy density of lithium ion batteries replaced other batteries in todays' consumer electronics and leading steps towards green energy. 2019 Nobel Prize in chemistry is awarded to Professor John Goodenough (The University of Texas at Austin, USA), M. Stanley Whittingham (Binghamton University, State University of New York, USA), and Akira Yoshino (Asahi Kasei Corporation, Tokyo, Japan, &Meijo University, Nagoya, Japan), for the development of lithium ion battery, Fig 2.

| 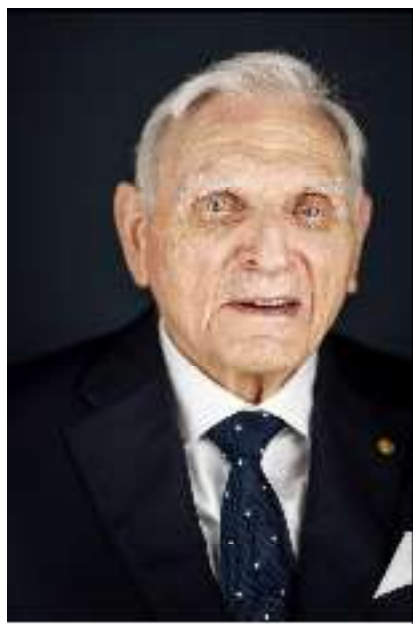 | 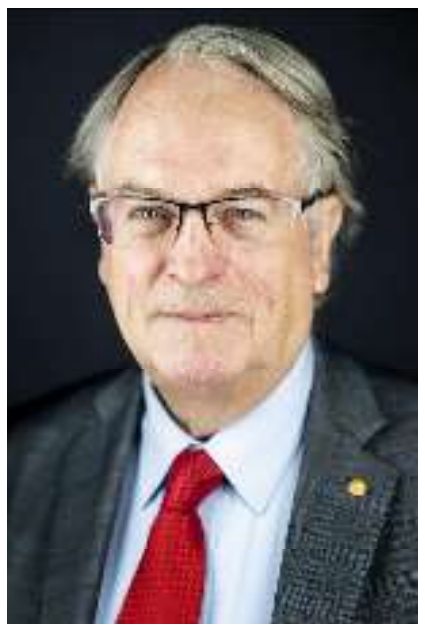 | 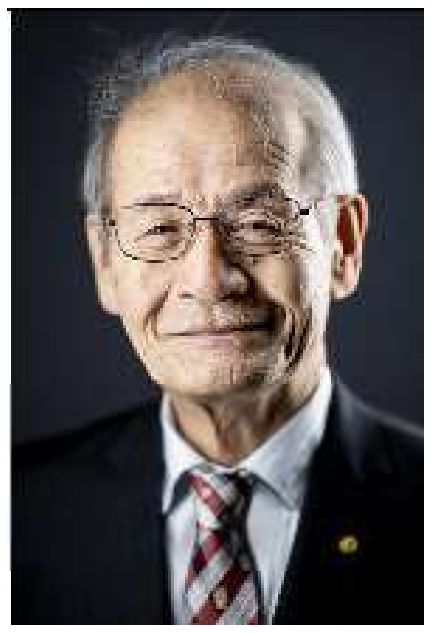 |
|---|---|---|
| **John B. Goodenough** The University of Texas at Austin, USA | **M. Stanley Whittingham** Binghamton University, State University of New York, USA | **Akira Yoshino** Asahi Kasei Corporation, Tokyo, Japan & Meijo University, Nagoya, Japan |
| Metal oxides as lithium intercalating cathodes e.g. Lithium cobalt oxide | Metal sulphide as lithium intercalating cathode e.g. titanium disulphide | Carbon material as an alternative anode material |

*Figure 2: Nobel prize winners and their contribution in the field of rechargeable energy storage, especially in lithium ion batteries,* **Reference: https://www.nobelprize.org/prizes/chemistry/2019/summary/**

The development of battery, especially lithium battery was initiated 1970 and accelerated onwards to explore efficient rechargeable electrical energy storage system, which can overcome the low gravimetric and volumetric energy density issues with contemporary rechargeable batteries. The work was fueled by these Nobel Prize winners' work, which revolutionized the field of electrical energy storage. Professor Whittingham, around 1970s, discovered titanium disulphide as a cathode material which can host and intercalate lithium ions efficiently in lithium ion batteries. Further, around 1980s, the work on metal oxide i.e. lithium cobalt oxide by Professor Goodenough provided an alternative cathode material to metal sulphide for lithium ion batteries which can operate at higher voltage ~.3.5 - 4V and thus, providing higher energy/power density. In these work metallic lithium was used as anode, which is highly reactive and explosive material, thus was not a common choice for battery applications. Later around 1985, Professor Akira Yoshino used a carbon material (petroleum coke) as an anode in place of lithium together with lithium cobalt oxide as cathode material to fabricate a lithium ion battery. This is considered as the birth point for the current commercially used lithium ion batteries. The advantage of such lithium ion batteries is that these are not based on any chemical reaction but intercalation of lithium ions via lithium ion movement between cathode and electrode during charging and discharging. The use of such rechargeable lithium ion batteries is increasing and most of the portable electronic devices are using only lithium ion batteries providing long hour power backups together with longer charge/discharge cyclability and thus, longer life span. With the advent of lithium ion batteries, there are continuous efforts in developing new battery materials, which can provide enhanced energy and power density, an essential requirement for power electronics and towards the development of electric and hybrid electric vehicles (cars, buses and even planes). However, there are still limited or negligible uses of lithium ion batteries for large power applications such as energy storage in solar photovoltaic power plants, hybrid electric vehicles i.e. plug-in electric vehicles and other such power applications. Further, large energy to power density aspect ratio together with higher operating voltage, large charge-discharge cyclability without capacity fading and safety issues are other important points to be considered towards developing such high energy and power density materials. Thus, large efforts are required to innovate such materials which can meet these requirements. The present article discusses the chronological development of cathode materials together with issues and challenges in realizing efficient rechargeable lithium ion batteries.

## Working Mechanism of a Lithium Ion Battery:

A battery stores electrical energy in the form of electrochemical energy, which can be converted back into electrical energy whenever required. It consists of four essential components (i) cathode (e.g. $LiFePO_4$), anode (e.g. nanocarbon based materials such as graphene), electrolyte (e.g. $LiPF_6$) a material allowing ions to travel through while electrically insulating avoiding electrical conduction through it, and a separator, a mesoporous material, allowing ions and also prevents electrical shorting between cathode and anode. All these components with representative materials are shown in Fig 3. The process is similar for other electrode materials. The respective lithium ion positions together with its relative motion are shown in Fig 3 for these charging, Fig 3(a), and discharging, Fig 3(b) states of a lithium ion battery. The insertion of lithium ion in cathode and anode is known as intercalation and mostly these cathode and anode materials are materials allowing easy intercalation rather chemical reaction. The charging and discharging are accompanied by lithium ion intercalation and deintercalation at the electrodesas illustrated below for $LiFePO_4$ cathode and carbon anode electrode materials.

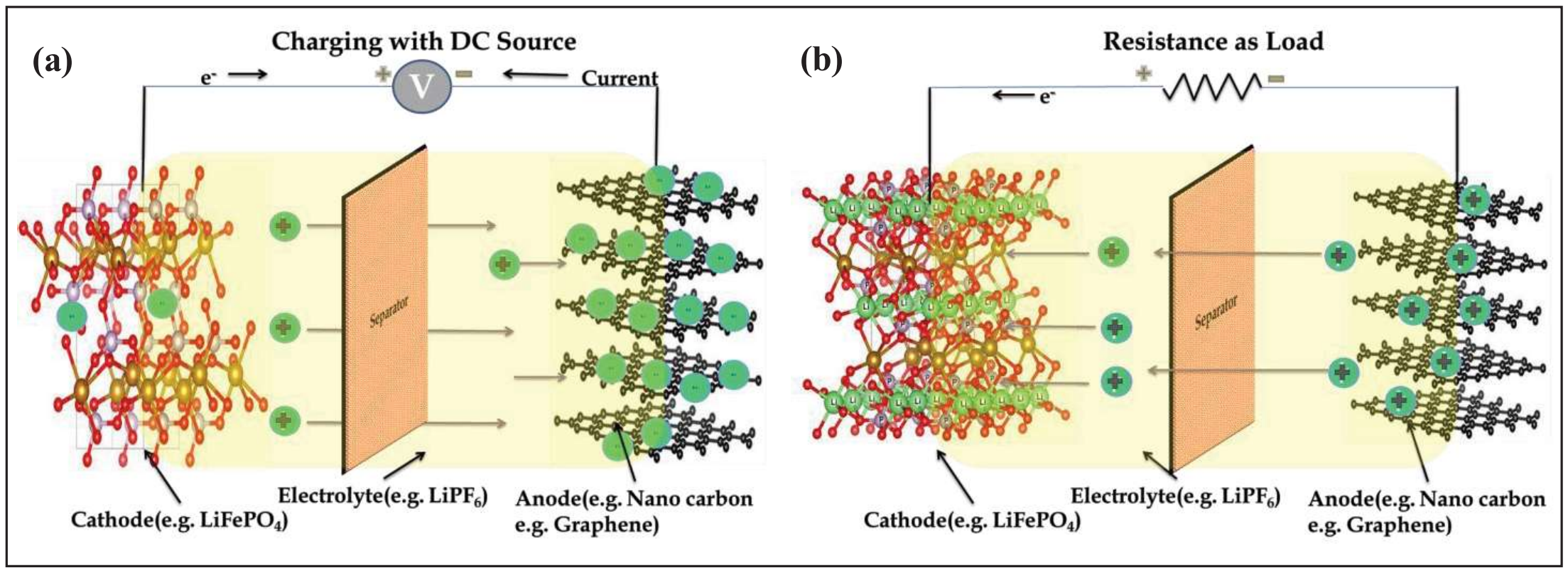

*Figure 3: Schematic representation of (a) charging using an external DC source and (b)discharging using an external load for a rechargeable Li-ion battery.*

At cathode or positive electrode

$$LiFePO_4 \underset{Discharging}{\overset{Charging}{\rightleftharpoons}} Li_{1-x}FePO_4 + x\,Li^+ + x\,e$$

At anode or negative electrode

$$C + x\,Li^+ + x\,e \underset{Discharging}{\overset{Charging}{\rightleftharpoons}} Li_xC$$

Total process = Sum of processes at cathode and anode

$$LiFePO_4 + C \underset{Discharging}{\overset{Charging}{\rightleftharpoons}} Li_{1-x}FePO_4 + Li_xC$$

The lithium ion forms complexes like LixC with carbon at anode ends during charging, whereas lithium atoms are at respective crystallographic sites in the cathode materials, which are mostly transition metals oxides, in the discharge states, Fig 3. The respective electron and current motions are shown in respective figures through load while discharging and through a voltage source while charging. The lithium chemical potential is lower in cathode than that of anode, which facilitates release of stored electrochemical energy into electrical energy while discharging. The most common electrolyte is $LiPF_6$ based organic liquid and stable upto 4.5 V and at higher voltage i.e. > 4.5 V, it may degrade or decompose into its constituents as well. Further, the operating voltage and energy density or battery capacity are limited by the electronic properties of cathode materials together with efficient ionic i.e. lithium ion transport. Thus, the development of suitable cathode material is very important not only to meet the high energy and power density requirements but also the operational safety.

## Configurations of rechargeable lithium ion batteries:

The increasing demand and compulsion towards green energy accelerated the development of rechargeable lithium ion batteries. Initially only smaller electronic systems/machineries were targeted and success can be noticed as lithium ion batteries replaced nearly completely its counterparts such nickel-cadmium and nickel-metal hydride rechargeable batteries. However, lithium ion batteries are still struggling to replace the conventional lead-acid batteries, which are mostly used in power sectors such as electric or hybrid-electric vehicles, power electronic equipment, and power plants. Considering the potential of lithium ion batteries and mitigating this gap, lithium ion batteries are developed in different configurations, as illustrated in Fig 4.

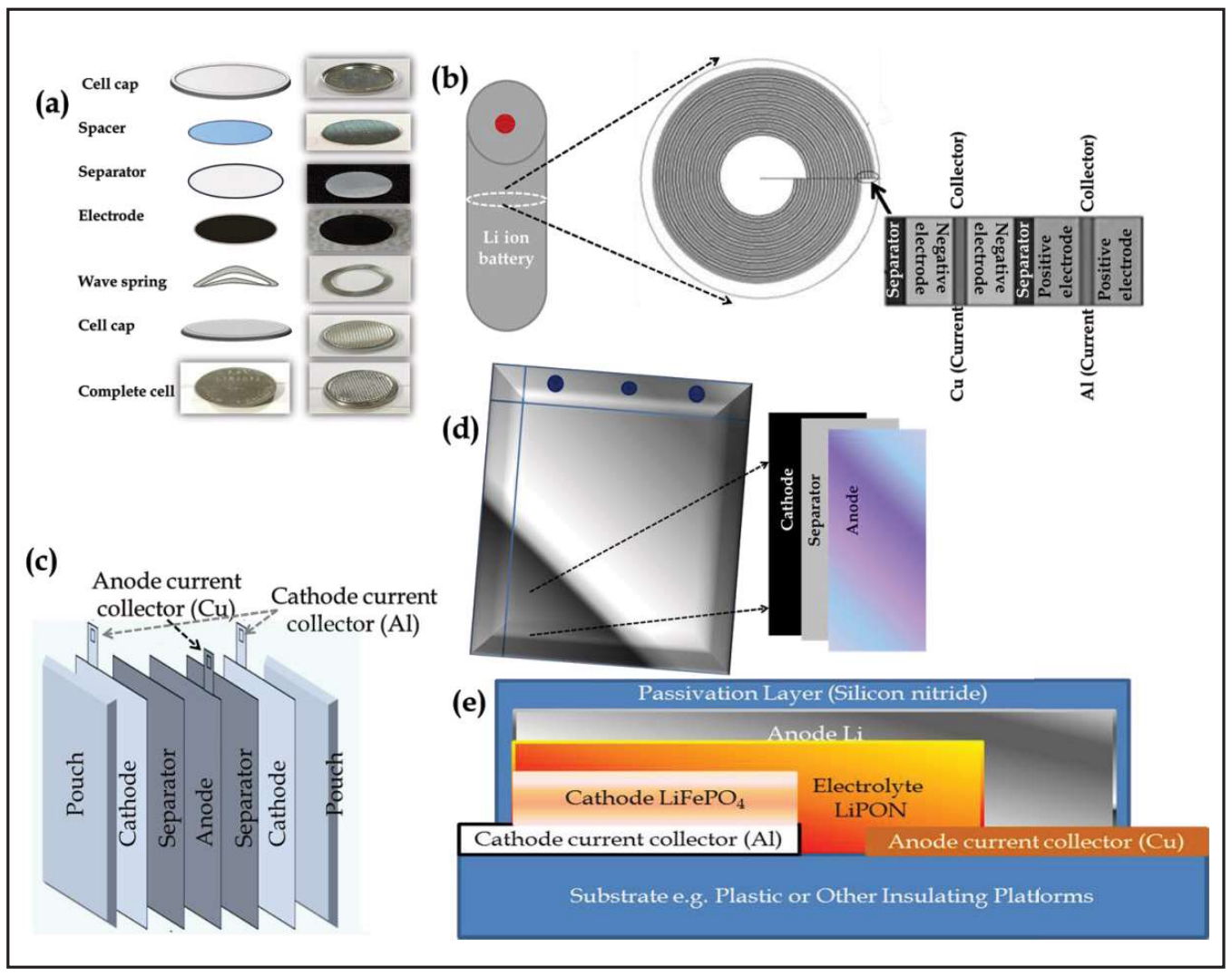

*Figure 4: The schematic diagrams for different configurations of lithium ion battery (a) coin or button cell, (b) cylindrical cell, (c) pouch cell, (d) prismatic cell, and (e) thin film cell*

The coin or button cell, Fig 4(a), is mostly used in laboratory to characterize the electrochemical performance of electrode materials. Here, each component of coin cell is shown schematically together with corresponding actual photographs from the authors' laboratory. The coin cell is also developed for commercial applications as well; however, it is mostly suitable for low power devices. The cylindrical cell is shown schematically in Fig 4(b)and used for powering devices with higher power requirements. These cells are combined in suitable (series and parallel) combinations to achieve the desired voltage and current. These coin and cylindrical cells use metallic casing and thus increasing overall weight of battery system. In this regard, pouch, Fig 4(c), may provide better solutions, as there is no metallic casing is used like coin cells. .Thus, these cells achieve the highest packing density and thus providing more energy density. The pouch cells are much larger than coin and cylindrical cells and used in various strategic applications such as space and military together with common electric and hybrid electric vehicles. Another similar cell is prismatic cell and shown schematically in Fig 4(d). These cells uses polymer and usually known as lithium-polymer cells. These pouch and prismatic cells may provide high power and very suitable for high current based applications. However, sometimes, swelling issues are noticed with pouch cells. Another interesting configuration is the thin film battery configuration. The geometrical configuration of thin film lithium ion battery is shown in Fig 4(e). This cell may offer a better solution to the present day power needs for miniaturized electronic devices and also probably may be integrated in the chip or devices itself.

## Development of cathode materials for lithium ion batteries and their classification:

The importance of lithium ion batteries compels to develop efficient batteries meeting the todays' electrical energy demand in all sectors including low power everyday applications to hybrid electrical vehicles and power plants. That's why there are continuous efforts in developing efficient materials to realize a miniaturized powerful battery together with operational safety. The inception of lithium ion battery in research can be noticed even in late 1950s with preliminary work on LiTMOx (TM = Co, Ni, Mn, Fe and V) and much emphasis was given on materials properties and their electrochemical performance. Further efforts ever made to develop new class of materials, as classified in Table 1.

**Table 1: Classification of cathode materials based on different properties and time period for the development of lithium ion batteries**

| S. No. | Generation | Cathode Materials | Operating Voltage range (V) | Capacity ($mAh\,g^{-1}$) | Time period |
|---|---|---|---|---|---|
| 1. | Generation I | Mono Li atom based transition metal (TM) ternary oxides | 1.2 – 3.5 | ≤ 150 | 1950 - 1985 |
| 2. | Generation II | Mono Li atom based TM ternary oxides | 3.5 – 4.0 | 150 – 200 | 1985 - 2000 |
| 3. | Generation III | Higher Li atom based TM ternary and quaternary oxides and oxifluorides | 3.5 – 5.0 | 200 – 330 or more | 2001 - 2020 |
| 4. | Generation IV | Li-S or Li-Air systems | 1.5 - 2.5 | >1000 | May be up to 2025 |

The development of new materials led to the enhanced operating voltage and energy density, Table 1, making lithium ion batteries suitable for various applications including electric vehicles and power plants. Further, pristine lithium ion based batteries are showing potential for very high energy density in both large and thin film geometries, Fig 1, however, large safety concerns are yet to address together with various technological issues before using in real applications. The next section will review the various materials developed over period with their salient materials and electronic features, suitable for lithium ion battery applications.

### 1st Generation materials for rechargeable lithium ion batteries (Since 1950- 1985):

The materials developed up to 1985 are classified as 1st generation cathode materials for rechargeable lithium ion batteries. These materials are summarized in Table 2. The crystallographic structures, their synthesis procedure adopted for fabricating such cathode materials are listed in Table 2 together with battery capacity and remarks, if any, with respective references.

This was the era to explore new potential cathode materials for rechargeable lithium ion batteries; that's why nearly all the transition metals are explored in the periodic table. As an essential condition, a multivalent ion based cathode material is required to compensate the charge during lithium insertion and extraction from the cathode material. The materials should show structural stability during lithium transport during charging/discharging process to realize large cycling stability. The manganese and vanadium based compounds show structural stability with relatively large specific capacity, which became the material of choice for next generation cathode materials. Most of these materials are transition metal based ternary oxides and mostly the materials and electrochemical properties are evaluated in their bulk forms. This is the reason of observing poor capacity with respect to their theoretical capacity at even much lower operating voltages. The main reason of lower capacity is their insulating nature together with relatively lower ionic conductivity and longer diffusion paths in bulk materials.

**Table 2: 1st Generation cathode materials used in lithium ion batteries (Time period 1950 – 1985)**

| S. No. | Material | Year | Configuration / structure | Method | Cathode/ Anode | Capacity ($mAhg^{-1}$) | Remark | Refs. |
|---|---|---|---|---|---|---|---|---|
| 1. | $Li_xCo_{1-x}O$ (x=.5) | 1958 | Rhombohedral | Solid state | Cathode | Not reported | First Commercial Li-ion battery material | 1 |
| 2. | $LiMnO_2$ | 1956 | Orthorhombic | Sintering | Cathode | Not reported | One of earliest material | 2 |
| 3. | $LiNiO_2$ | 1958 | Rhombohedral | Solid State | Cathode | Not reported | None | 3 |
| 4. | $Li_xCoO_2$ | 1980 | Hexagonal | Electrochemical extraction | Cathode | 4.7 Volt at X=0.1 | None | 4 |
| 5. | $Li_{1-x}MnO_2$ | 1983 | Tetragonal | Lithiation of $\beta$-$MnO_2$ | Cathode | Not Reported | None | 5 |
| 6. | $Li_{1.5}Fe_3O_4$ | 1982 | Cubic | Chemical & electrochemical | Cathode | 1.3 Volt | None | 6 |
| 7. | $Li_{1.7}Fe_2O_3$ | 1982 | Cubic | Chemical & electrochemical | Cathode | 1.2 Volt | None | 6 |
| 8. | $LiMn_2O_4$ | 1983 | Cubic | Solid state | Cathode | Not reported | None | 7 |
| 9. | $Li_2Mn_2O_4$ | 1983 | Tetragonal | Solid state | Cathode | Not reported | None | 7 |
| 10. | $LiMn_3O_4$ | 1983 | Tetragonal | Solid state | Cathode | Not reported | None | 7 |
| 11. | $Li_{1-x}V_2O_4$ (0<x<.34) | 1985 | Cubic | Solid state | Cathode | Not reported | None | 8 |
| 12. | $Li_{1.5}V_2O_4$ | 1985 | Cubic | Solid state | Cathode | Not reported | None | 8 |
| 13. | $Li_2V_2O_4$ | 1985 | Cubic | Solid state | Cathode | Not reported | None | 8 |

## 2nd Generation materials for rechargeable lithium ion batteries (Since 1986- 2000):

The introduction of transition metal oxides based cathode materials by Professor J. B. Goodenough and his team already showed the potential of lithium ion batteries in 1st generation. However, the limited capacity and lower operating voltages compelled to innovate the modification strategies of 1st generation cathode materials or design and develop new materials for desired electrochemical properties to achieve the optimal performance. The issues with the first generation materials were their poor electrical and ionic conductivity and efforts were put to mitigate these issues in second generation. The doping strategies were adopted to modify the ionic conductivity together with electronic conductivity. The most of ternary oxides from 1st generations are followed in second generation with suitable dopants. The doping of high valent either transition metals or non-transition metals at second cation sites led to enhanced electronic conductivities of these materials. The 2nd generation cathode materials are summarized in Table 2 together with their synthesis process, crystallographic phases, electrochemical capacity together with some noticeable properties as remarks.

Further, in contrast to the bulk materials, nanostructured e.g. nanoparticle, nanotubes, nanorods or two dimensional (2D) sheets like structures are synthesized for these materials and used for electrochemical studies. The reduced dimensions of these materials showed improved ionic conductivities as it resulted in the reduced diffusion paths for lithium ions in case of nanomaterials with respect to bulk materials. The electronic conductivities of these materials are also improved by designing cathode material as core and graphitic carbon as shell, leading to several orders of magnitude higher electrical conductivity. In addition to the first generation cathode materials, emphasis was also given to the vanadium based ternary oxide materials as potential cathode materials, Table 2, due to their layered structures having large potential for easy intercalation and deintercalation, showing moderated ~ 150 mAh g$^{-1}$ capacity. The structural instability associated with these materials for complete removal of lithium hampered the use of full capacity of these cathode materials. Most interestingly, again from Professor J. B. Goodenough's group, a quaternary oxide $LiFePO_4$ as cathode material was reported during 1997, as a potential candidate in its olivine structure and showed electrochemical capacity, very close to the theoretical limits, 170 mAh g$^{-1}$, after modifying its electronic and ionic conductivities. Olivine $LiFePO_4$ system among its derivatives with other crystallographic structures is very robust, showing no crystallographic degradation even after complete deintercalation of lithium and showed very high stability. Thus, $LiFePO_4$ can exhibit very large charge/discharge cycles without any capacity fading.

**Table 3: 2nd Generation cathode materials used in lithium ion batteries (Time period 1986 – 2000)**

| S. No. | Material | Year | Configuration / structure | Method | Cathode/ Anode | Capacity ($mAhg^{-1}$) | Remark | Refs. |
|---|---|---|---|---|---|---|---|---|
| 1. | $Li(Co_{1-x}Li_{x/3}Mn_{2x/3})O_2$(x=.0) x = 0.1, 0.2, 0.3 | 1999 | Monoclinic | Solid state | Cathode | 160$mAhg^{-1}$ | Stable | 9 |
| 2. | $LiCoO_2$, $LiNiO_2$ | 1990 | Hexagonal | Solid state | Cathode | Not reported | None | 10 |
| 3. | LT-$LiCoO_2$ [Layered] (low temperature) | 1993 | Trigonal, Cubic | Nitrate method, Carbonate method | Cathode | Not reported | Prepared at low temp. | 11 |
| 4. | $LiCoO_2$ | 1996 | Cubic | Microwave heating | Cathode | 140 $mAhg^{-1}$ | Stable and good cyclability | 12, 13 |
| 5. | $LiCoO_2$ | 1996 | Trigonal | Sol-gel | Cathode | 98 $mAhg^{-1}$ | None | 14 |
| 6. | $LiCoO_2$ (Pseudo spinel) | 1996 | Cubic | Solid state | Cathode | Not reported | None | 14 |
| 7. | $LiCoO_2$[Layered] | 1998 | Trigonal | Solid state | Cathode | 150$mAhg^{-1}$ | None | 15 |
| 8. | $LiCoO_2$ | 1998 | Hexagonal | Solid state | Cathode | 142$mAhg^{-1}$ | Stable, good rate capability & good cyclability | 16 |
| 9. | $LiCoO_2$ | 1998 | Hexagonal, Trigonal | Solid state, Sol-gel | Cathode | 145$mAhg^{-1}$ -165$mAhg^{-1}$ | None | 17, 18, 19 |
| 10. | $LiCoO_2$ | 1999 | Hexagonal | Sol-gel | Cathode | 120$mAhg^{-1}$ | None | 20, 21 |
| 11. | Corrugated layer $LiFeO_2$ | 1996 | Orthorhombic | Sol-gel | Cathode | 110mAhg$^{-1}$ | Synthesized at 90 °C | 22 |
| 12. | Goethite type $LiFeO_2$ | 1996 | Orthorhombic | Sol-gel | Cathode | 65mAhg$^{-1}$ | Synthesized at 170 °C | 22 |
| 13. | $Li_3Fe_2(PO_4)_3$ | 1997 | Monoclinic | Solid state | Cathode | 105 mAhg$^{-1}$ | None | 23 |
| 14. | $LiFeP_2O7$ | 1997 | Monoclinic | Solid state | Cathode | 65 mAhg$^{-1}$ | None | 23 |
| 15. | $LiFePO_4$ | 1997 | Hexagonal | Solid state | Cathode | 130 mAhg$^{-1}$ | High capacity | 23, 24 |
| 16. | $LiFePO_4$ | 1999 | Orthorhombic | Solid state | Cathode | 125 mAhg$^{-1}$ | None | 25 |
| 17. | $LiMn_2O_4$ | 1991 | Trigonal | Solid state | Cathode | Not reported | None | 26 |
| 18. | $LiNiO_2$ | 1991 | Rhombohedral | Solid state | Cathode | Not reported | None | 26 |
| 19. | $LiCoO_2$ | 1991 | Trigonal | Solid state | Cathode | Not reported | None | 26 |
| 20. | $LiMn_2O_4$ | 1996 | Cubic | Solid state | Cathode | 75mAhg$^{-1}$ | None | 27, 28 |
| 21. | $LiMnO_2$ | 1996 | Monoclinic | Solid state | Cathode | 280mAhg$^{-1}$ | Pure material, High capacity & less fading | 29 |
| 22. | $LiNiO_2$ | 1990 | Hexagonal | Solid state | Cathode | .009Ahg$^{-1}$ | Long cycle life | 30,31 |
| 23. | $LiNiO_2$ | 1995 | Hexagonal | Solid state | Cathode | 260mAhg$^{-1}$ | High capacity | 32 |
| 24. | $TiO_2$ ($LiCoO_2$ as cathode) | 1995 | Tetragonal | As prepared | Anode | 46 mAhg$^{-1}$ | None | 33 |
| 25. | $LiV_2O_4$ | 1986 | Trigonal | Solid state | Cathode | Not reported | None | 34 |
| 26. | $LiVO_2$ | 1988 | Cubic | Solid state | Cathode | Not reported | None | 35 |
| 27. | $Li_2MnO_2$ | 1990 | Trigonal, Hexagoanl | Solid state | Cathode | Not reported | Changes in phase by Li insertion | 36 |
| 28. | $LiV_2O_4$ | 1991-99 | Cubic | Solid state | Cathode | .117Ahg-$^{1}$, 135AhKg$^{-1}$ | Initial structure retained after 500 cycle,70% retention after 800 cycle | 35, 37, 38, 39, 40 |

The second generation materials showed guidelines to modify the ionic and electronic properties of cathode materials showing improved performance. However, the power density was still limited not only in terms of energy density but also in terms of operating voltage and current. The operating voltage of 2$^{nd}$ generation materials is limited to 3.5 V, whereas there are some studies showing high current extraction using nanostructured cathode materials in lithium ion batteries.

**3rd Generation materials for rechargeable lithium ion batteries (Since 2001- 2020):**

The initially developed cathode materials showed their potential in the development of commercial lithium ion batteries, and are widely used today for powering everyday electronics in general together with applications in strategic areas such as space and defence. However, limited energy density and relatively lower operating voltage motivated to explore new cathode materials, which may probably mitigate such issues. In third generation, emphasis was given to develop cathode materials with very high electrochemical capacity ( 250mAh $g^{-1}$ or more) with good charge/discharge cyclability. These requirements led to the surge in the development of cathode materials, as can be seen in Table 4, listing the developed cathode materials during this period. A large number of cathode materials are reported in table 4, showing relatively higher electrochemical capacities.

**Table 4: 3rd Generation cathode materials used in lithium ion batteries (Time period 2001 – 2020)**

| S. No. | Material | Year | Configuration /structure | Method | Cathode/ Anode | Capacity ($mAhg^{-1}$) | Remark | Refs. |
|---|---|---|---|---|---|---|---|---|
| 1. | $LiVPO_4OH$ | 2016 | Monoclinic, Triclinic | Hydrothermal | Cathode | 280$mAhg^{-1}$ | New tavorite type composition | 41 |
| 3. | $LiVPO_4F$ | 2003, 2014 | Triclinic | Carbothermal reduction | Cathode | 116$mAhg^{-1}$ | None | 42 - 44 |
| 4. | $LiVPO_4F$ (Ti doped) | 2014 | Triclinic | Carbothermal reduction | Cathode | 128 $mAhg^{-1}$ | None | 43,44 |
| 5. | $LiVPO_4F$ ,$LiVPO_4F$ /Ag | 2015 | Triclinic | Sol-gel | Cathode | 102$mAhg^{-1}$ - 117$mAhg^{-1}$ | None | 45 |
| 6. | $LiVPO_4F$/ C,$LiVPO_4F$/C-N | 2016, 2019 | Triclinic | Sol-gel | Cathode | 140$mAhg^{-1}$ | None | 46, 47 |
| 7. | $Li_4Ti_5O_{12}$ | 2007 | Cubic | Sol-gel | Anode | 170$mAhg^{-1}$ | None | 48 |
| 8. | $Li[Cr_xLi_{(1-x)}Mn_{2(1-x)}]O_2$ | 2007 | Hexagonal | Sol-gel | Cathode | 195 $mAhg^{-1}$ | Stable cycling performance at x=.290 | 49 |
| 9. | $LiVPO_4F$, $LiY_xV_{(1-x)}$ PO4F(x=0.04) | 2009 | Triclinic | Carbothermal reduction | Cathode | 119 $mAhg^{-1}$ - 134 $mAhg^{-1}$ | Stable cyclic performance | 50 |
| 10. | $LiVPO_4F$ | 2010 | Triclinic | Sol-gel | Cathode | 134 $mAhg^{-1}$ | None | 51 |
| 11. | $Li_4Ti_5O_{12}Li_4Ti_5O_{12}$/ Sn | 2011 | Cubic | Solid-state | Anode | 165 – 321 $mAhg^{-1}$ | None | 52 |
| 12. | $LiVPO_4F$/grphene | 2017 | Triclinic | Ionothermal process | Cathode | 129 - 153$mAhg^{-1}$ | Good cycling stability and good discharge efficiency | 53 |
| 13. | $LiVPO_4F$/C | 2018 | Triclinic | Sol-gel | Cathode | 135.3$mAhg^{-1}$ | Stability in cyclability | 54 |
| 14. | $LiVPO_4F$/C | 2018 | Triclinic | Sol-gel | Cathode | 140$mAhg^{-1}$ | None | 55 |
| 15. | $LiV_2O_4$ | 2004 | Cubic | Solid state | Cathode | Not reported | None | 56 |
| 16. | $LiNi_{0.5}Co_{0.2}Mn_{0.3}O_2$, $LiV_2O_4$ coated $LiNi_{0.5}Co_{0.2}Mn_{0.3}O_2$ | 2019 | Hexagonal | Solid-state | Cathode | 145 - 164$mAhg^{-1}$ | None | 57 |
| 17. | $LiNi_{1-x}Sb_xO_2$(x=0, 0.1, 0.15, 0.2) | 2011 | Hexagonal | Solid-state | Cathode | 105 -117$mAhg^{-1}$ | Structure stability and good cyclability | 58 |
| 18. | $LiMnPO_4$/C | 2011 | Orthorhombic | Spray pyrolysis | Cathode | 149 $mAhg^{-1}$ | None | 59 |
| 19. | $Li[Li_{0.2}Mn_{0.54}Ni_{0.13}Co_{0.13}]O_2$ | 2011 | Trigonal | Co-precipitation | Cathode | 250 - 268 $mAhg^{-1}$ | None | 60 |
| 20. | $Li_{1.2}Mn_{.56}Ni_{.16}Co_{.08}O_2$ | 2014 | Monoclinic & Rhombohedral | Solid-state | Cathode | 245 $mAhg^{-1}$ | Good cyclability | 61 |
| 21. | $Li[Li_{0.2}Mn_{0.54}Ni_{0.13}Co_{0.13}]O_2$ (nanoplate) | 2016 | Hexagonal | Co-precipitation | Cathode | 307.9 $mAhg^{-1}$ | Excellent cyclability | 62 |
| 22. | $Li_{1.1}Ni_{.35}Mn_{.65}O_2$ | 2017 | Hexagonal | Solid-state | Cathode | 210 $mAhg^{-1}$ | None | 63 |
| 23. | $Li_{1.2}Mn_{.6}Ni_{.2}O_2$-SG | 2018 | Hexagonal | Sol-gel | Cathode | 230.38 $mAhg^{-1}$ | None | 64 |
| 24. | $Li_{1.17}Mn_{0.56}Ni_{0.14}Co_{0.13}O_2$ (PEG2000-LMNCO) | 2015 | Hexagonal | Sol-gel | Cathode | 228 $mAhg^{-1}$ | None | 65 |
| 25. | $LiCoO_2$($Li_4Ti_5O_{12}$ coated) | 2007 | Hexagonal | Sol-gel | Cathode | 179 $mAhg^{-1}$ | None | 66 |
| 26. | $Li_4Ti_5O_{12}$ | 2008 | Cubic | Solid-state | Cathode | 170 $mAhg^{-1}$ | None | 67 |
| 27. | $Li_4Ti_5O_{12}/B_0$-C, $Li_4Ti_5O_{12}/B_{0.1}$-C, $Li_4Ti_5O_{12}/B_{0.3}$-C,$Li_4Ti_5O_{12}/B_{0.5}$-C | 2016 | Cubic | Solid-state | Cathode | 154 $mAhg^{-1}$ | Good cyclability | 68 |
| 28. | $LiCoO_2$(Pristine and $Al_2O_3$ doped) | 2017 | Hexagonal | Solid-state | Cathode | 195 – 202 $mAhg^{-1}$ | None | 69 |
| 29. | $LiCoPO_4$ | 2010 | Orthorhombic | Solid-state Rheological phase method | Cathode | 30.9-71.5 $mAhg^{-1}$ | None | 70 |
| 30. | $Li_2MnSiO_4$ /Pure& carbon coated | 2012 | Orthorhombic | Sol-gel | Cathode | 81 – 144 $mAhg^{-1}$ | None | 71 |
| 31. | $Li_2Mn_{.80}Fe_{.20}SiO_4$ ($SiO_2$ size 5nm) | 2013 | Orthorhombic | Hydrothermal | Cathode | 95.6 – 129.4 $mAhg^{-1}$ | None | 72 |
| 32. | $Li_2MnSiO_4$ /C, $Li_2Mn_{.96}Ca_{.04}SiO_4$ /C | 2018 | Orthorhombic | Solvothermal | Cathode | 141.1 $mAhg^{-1}$ | None | 73 |
| 33. | $Li_2MnSiO_4$ / C$Li_2MnSi_{.75}V_{.25}O_4$ /C | 2016 | Orthorhombic | Sol-gel | Cathode | 60 – 130 $mAhg^{-1}$ | None | 74 |
| 34. | B-$Li_2MnSiO_4$(Bulk & porous LMS) | 2015 | Orthorhombic | Hydrothermal | Cathode | 120 - 217$mAhg^{-1}$ | None | 75 |
| 35. | $Li_2MnSiO_4$/C (20AM+3KB+3TAB) AM-Active Material KB-Ketjen Black TAB-Teflonized Acetylene Black | 2011 | Orthorhombic | Solid State | Cathode | 79 – 160 $mAhg^{-1}$ | Good cyclability | 76 |
| 36. | $Li_2MnSiO_4$ Powder | 2012 | Orthorhombic | Sol-gel | Cathode | 206 - 295 $mAhg^{-1}$ | Good cyclability | 77 |
| 37. | $Li_2MnSiO_4$and derivative systems | 2013 | Orthorhombic | Hydrothermal | Cathode | 185 $mAhg^{-1}$ | None | 78 |
| 38. | $Li_2MnSiO_4$/C and derivative systems | 2016 | Orthorhombic | Hydrothermal | Cathode | 135 $mAhg^{-1}$ | None | 79 |
| 39. | $Li_2MnSiO_4$& Mg, Al, Ga doped sample | 2013 | Orthorhombic | Sol-gel | Cathode | 101 – 147 $mAhg^{-1}$ | None | 80 |
| 40. | $Li_2MnSiO_4$/C nano Composites | 2010 | Monoclinic | Microwave-solvothermal | Cathode | 215 $mAhg^{-1}$ | None | 81 |
| 41. | $Li_2MnSiO_4$ ($Pmn2_1$) | 2011 | Orthorhombic | Sol-gel | Cathode | 100 $mAhg^{-1}$ | None | 82 |
| 42. | $Li_2MnSiO_4$ as Cathode | 2016 | Orthorhombic | Sol-gel | Cathode | 405 $mAhg^{-1}$ | None | 83 |
| 43. | $Li_2MnSiO_4$ as Anode | 2016 | Orthorhombic | Sol-gel | Cathode | 658 $mAhg^{-1}$ | High energy density and good cyclability | 83 |
| 4. | $Li_2MnSiO_4$ (.5 hwr ball milled) | 2007 | Orthorhombic | Sol-gel | Cathode | 142 $mAhg^{-1}$ | None | 84 |
| 45. | $Li_2MnSiO_{3.97}F_{0.03}$ (X=0.03) | 2018 | Orthorhombic | Solid-state | Cathode | 279 $mAhg^{-1}$ | Stable energy density | 85 |
| 46. | $Li_2MnSiO_4$(at 55° C) | 2014 | Orthorhombic | Sol-gel | Cathode | 225 $mAhg^{-1}$ | None | 86 |
| 47. | $Li_2MnSiO_4$(at 700° C) | 2014 | Orthorhombic | Molten salt | Cathode | 165 $mAhg^{-1}$ | Good cyclability | 87 |
| 48. | $Li_2MnSiO_4$/C | 2016 | Orthorhombic | Sonochemic-al reaction | Cathode | 261 $mAhg^{-1}$ | None | 88 |
| 49. | $Li_2MnSiO_4$/ C(Carbon wt% 2.1) | 2013 | Orthorhombic | Sol-gel | Cathode | 145 $mAhg^{-1}$ | None | 89 |
| 50. | $Li_2MnSiO_4$-Pristine | 2016 | Orthorhombic | Sol-gel | Cathode | 203 $mAhg^{-1}$ | None | 90 |
| 51. | $Li_2FeSiO_4$ | 2013 | Monoclinic | Sol-gel | Cathode | 180 $mAhg^{-1}$ | Stable structure and good cyclability | 91 |
| 52. | $Li_2Mn_{0.94}Mo_{.06}SiO_4$ | 2015 | Orthorhombic | Sol-gel | Cathode | 207.3 $mAhg^{-1}$ | None | 92 |
| 53. | Amorphous type $Li_2Mn_{0.85}Ti_{0.15}SiO_4$/ C(S-LMST) | 2018 | Orthorhombic | Hydrothermal | Cathode | 185 $mAhg^{-1}$ | None | 93 |
| 54. | $Li_2Mn_{.9}Ti_{0.1}$SiO4 | 2015 | Orthorhombic | Sol-gel | Cathode | 211 $mAhg^{-1}$ | None | 94 |
| 55. | $Li_2MnSiO_4$/C with Ni doping | 2014 | Orthorhombic | Solvothermal | Cathode | 274.5 $mAhg^{-1}$ | High energy density | 95 |
| 56. | $Li_2Mn_{.8}Fe_{.2}SiO_4$ | 2015 | Orthorhombic | Sol-gel | Cathode | 122 $mAhg^{-1}$ | None | 96 |
| 57. | $Li_2Mn_{.9}Cu_{.1}SiO_4$ | 2015 | Orthorhombic | Sol-gel | Cathode | 206 $mAhg^{-1}$ | None | 97 |
| 58. | CNT@$Li_2MnSiO_4$@C | 2018 | Orthorhombic | Sol-gel | Cathode | 227 $mAhg^{-1}$ | None | 98 |
| 58. | MP-$Li_2MnSiO_4$@C (MP- mesoporous morpholosy) | 2018 | Orthorhombic | *In situ* Template method based on non ionic surfactant F127 | Cathode | 164 $mAhg^{-1}$ | Good cyclability | 99 |
| 59. | $Li_2Mn_{.99}La_{.01}SiO_4$ | 2015 | Orthorhombic | Hydrothermal | Cathode | 257$mAhg^{-1}$ (at.05C) | None | 100 |
| 60. | $Li_2Mn_{.925}Cr_{.075}SiO_4$ | 2018 | Orthorhombic | Solid-State | Cathode | 200 $mAhg^{-1}$ | None | 101 |
| 61. | $Li_2$(Mn/Fe)SiO4 | 2007 | Orthorhombic | Sol-gel | Cathode | 125 $mAhg^{-1}$ | None | 102 |
| 62. | $Li_{1.95}Mn_{.95}Cr_{.05}SiO_4$ | 2014 | Orthorhombic | Sol-gel | Cathode | 238 $mAhg^{-1}$ | High energy density | 103 |
| 63. | $Li_{2.05}Mn_{.95}Al_{.05}SiO_4$ | 2014 | Orthorhombic | Sol-gel | Cathode | 220 $mAhg^{-1}$ | High energy density | 103 |
| 64. | $Li_2MnSiO_4$/NC-2 | 2019 | Orthorhombic | Sol-gel | Cathode | 276.88 $mAhg^{-1}$ | High energy density & good cyclability | 104 |
| 65. | $Li_2MnSiO_4Li_2FeSiO_4$ | 2012 | Orthorhombic | A rapid one-pot supercritical fluid reaction | Cathode | 350 - 320 $mAhg^{-1}$ | High energy density and excellent cyclability | 105 |
| 66. | Annealed $Li_2MnSiO_4$ (A-LMS) | 2012 | Orthorhombic | Hydrothermal | Cathode | 226 $mAhg^{-1}$ | High energy density | 106 |
| 67. | $Li_2MnSiO_4$/C-2 (Prepa-red in $(BMIM)BF_4$) | 2014 | Orthorhombic | Ionothermal | Cathode | 218 $mAhg^{-1}$ | Good cyclability | 107 |
| 68. | $Li_2MnSiO_4$/CNFs | 2015 | Orthorhombic | Solvothermal | Cathode | 350 $mAhg^{-1}$ | High energy density & good cyclability | 108 |
| 69. | $Li/Li_{1.8}MnSi_{0.8}P_{0.2}O_4$ | 2014 | Orthorhombic | Sol-gel | Cathode | 155 $mAhg^{-1}$ | None | 109 |
| 70. | $Li_2MnSiO_4$/C | 2014 | Orthorhombic | Sol-gel | Cathode | 256.86 $mAhg^{-1}$ | None | 110 |
| 71. | $Li_2Mg_{.1}Mn_{.9}SiO_4$/C | 2011 | Orthorhombic | Sol-gel | Cathode | 289 $mAhg^{-1}$ | High energy density & low energy density | 111 |
| 72. | $Li_2MnSiO_4$ | 2011 on-wards | Orthorhombic | Solid-State, sol-gel, hydrothermal | Cathode | < 100– 268 $mAhg^{-1}$ | poor retention, and unclear cyclic stability | 112-126 |

The 3rd generation materials are also transition metal quaternary oxides and oxyfluorides, Table 4. The vanadium based $LiVPO_4$ and its derivatives showed relatively higher capacities (> 250 mAh $g^{-1}$). Further, to increase electrochemical storage capacity, emphasis was given to develop materials with higher lithium content with respect to mono-lithium based cathode materials. The theoretical capacity of higher lithium content such as $Li_2TMSiO_4$ (TM = transition metal e.g. Fe, Co, Ni, Mn, V etc) is around 333 mAh $g^{-1}$, nearly twice to that of mono-lithium based cathode materials. Further, their large band gap may lead to higher operating voltage as well. These materials are relatively new cathode materials and explorations are still going on. The initial studies showed high electrochemical capacity for such compounds, which starts fading after first complete charge/discharge cycles and showed reduction to very low 50-60 mAh $g^{-1}$ in 10 – 20 charge/discharge cycles. This is attributed to the structural instability of $Li_2TMSiO_4$ cathode materials. The initial structure is subjected to Jahn-Teller (J-T) distortion during charging/discharging and may lead to permanent crystallographic changes at local levels, causing poor cyclability. However, further studies are needed to understand this degradation mechanism and strategies need to be evolved to overcome this aspect. The nanostructuring and even localized strain by doping may provide some understanding towards mitigating the capacity fading in these cathode materials, which seems to potential candidate for high energy density cathode materials.

## 4thGeneration materials for rechargeable lithium ion batteries (2020 onwards):

The current developments (up to 3rd generations) in rechargeable are widely suitable and used for low power applications in our everyday life and mostly limited with their specific current density up to 150 – 200 mAh g-1. The scaling of these batteries is realized in terms of large battery packs for specific purposes including electric and hybrid- electric vehicles. The power hungry applications such as electric and hybrid electric vehicles, planes, and other commuting means toward green energy initiatives may need very high energy density (> 1000 mAh $g^{-1}$ or more).

The problem of limited specific capacity can be taken care by innovating new materials and battery designs which can show very high capacity i.e. > 1000 mAh g-1. For example metal air batteries are showing promise to beat the current specific capacity limitations, a hurdle in the present lithium ion batteries. The lithium- air and also lithium – sulfur batteries have shown promise with very high specific energy densities. However, the research and development of such batteries are still in nascent stages and various issues such as safety because of the use of pure lithium, which is highly flammable and cyclability need to address carefully.

## Hunt for new lithium ion battery cathode materials and design criteria:

The performance of a lithium battery depends on operating voltage and current, energy density, cyclability, stability, and most importantly safety. The electronic properties of cathode materials such as band gap and relative electronic state positions affect the operating voltage of a lithium ion battery. The relative positions of electronic states in terms of electrochemical potential of respective electrodes are shown in Fig 5 for a lithium ion battery. The electrolyte is a good ionic conductor but electrically insulator, with a large band gap $E_g$ expressed as the difference between the lowest unoccupied molecular orbital (LUMO or also known as conduction band) and the highest occupied molecular orbital (HOMO or valence band), as shown schematically in Fig 5. This band gap of electrolyte is also known as the maximum allowed voltage window as above or below these energy levels, the electrolyte will start dissociating.

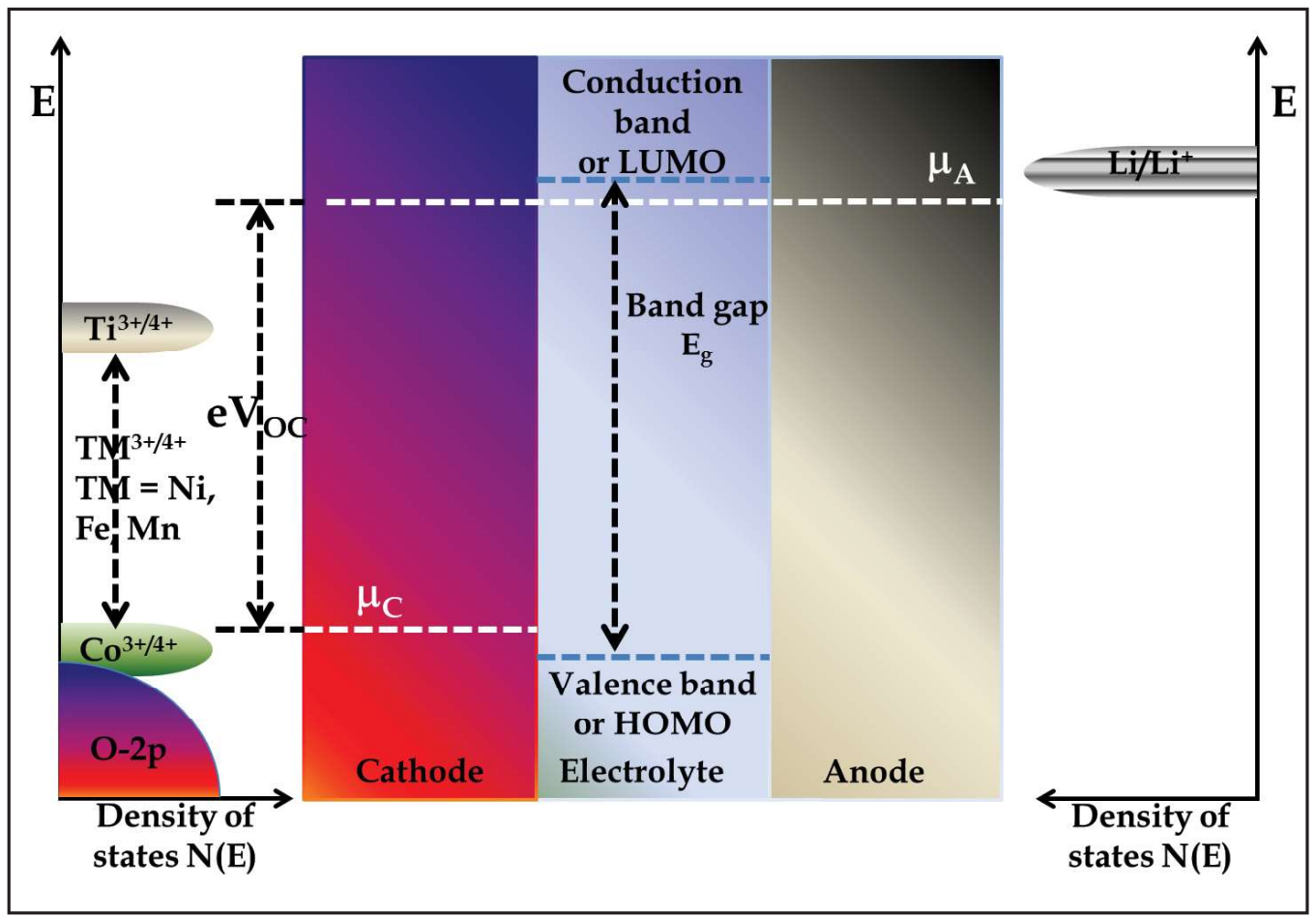

*Figure 5: The schematic diagram of electronic states in terms of chemical potential positions of cathode, anode together with band gap of electrolyte. The left most figure shows the relative position of different transition metal elements with respect to oxygen 2p states in transition metal oxides. The right most figure shows the Fermi energy of lithium metal.*

Further, the relative chemical potential of cathode and anode are shown with respect to electrolyte HOMO and LUMO positions, suggesting that the open circuit voltage $eV_{OC} = \mu_A - \mu_C$, where e is electronic charge (=1.6x10$^{-19}$ C), $V_{OC}$ is open circuit voltage, $\mu_A$ and $\mu_C$ are chemical potential of anode and cathode materials, respectively. Thus, the cathode material should be selected in such a way that it's chemical potential i.e. $\mu_C$ is located above the valence band (i.e. HOMO) of electrolyte, else the electrolyte will oxidize at the cathode-electrolyte interface, leading to a thin film formation at solid-electrolyte interface and is commonly known as SEI film. On the other hand, the chemical potential of anode i.e. $\mu_A$ is located below the conduction band (i.e. LUMO) of the electrolyte, else electrolyte will reduce at the anode-electrolyte interface, causing SEI formation at this interface.Thus, the selection of cathode and anode is limited by the electrochemical window i.e. band gap and relative position of HOMO and LUMO of electrolyte. However, there are continuous efforts to increase the operating voltage of lithium ion batteries and either existing materials are engineered or new materials are being developed to meet the requirements of higher operating voltages together with enhanced electrochemical energy storage capacity. Numerous materials are investigated, as listed in Table 4, where high lithium content materials have shown promise with respect to mono-lithium based cathode materials. However, there are various technological issues and challenges such as their capacity fading in few cycles. Thus, efforts are required to address the capacity fading in high lithium based cathode materials to increase the electrochemical energy density. However, there is a need to develop not only cathode materials with higher operating voltage and enhanced energy density but also compatible electrolyte simultaneously.

## Conclusion:

This review addresses the development of cathode materials for rechargeable lithium ion batteries. Large efforts are put to realize the commercial lithium ion batteries, however, there are still issues and challenges to make it more suitable for large scale power applications such as power plants, hybrid electric vehicles. The three generations of cathode materials are discussed covering from its inception to the current state of materials. The large emphasis is given to enhance the electrochemical energy storage density and operating voltage of lithium ion batteries to meet the future demands in all sectors. The associated crystallographic instability in $Li_2TMSiO_4$ based cathode materials is hampering the full potential of these cathode materials. Further, various layered, spinel, and olivine structured materials are widely explored and also shown potential for future energy needs. However, the success of high voltage and high electrochemical energy density based cathode materials will rely on simultaneous development of suitable electrolytes.

## Acknowledgement:

Author acknowledges Mr. Kuldeep Kaswan for his assistance in compiling cathode materials and also funding agency Department of Science and Technology (DST), Government of India (project no. INT/RUS/RFBR/320).

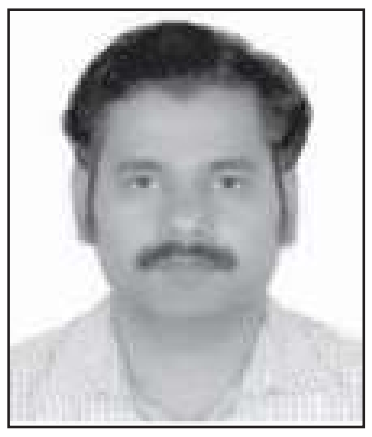

**Dr. Ambesh Dixit** *is an Associate Professor at the Indian Institute of Technology Jodhpur (IIT Jodhpur), India, and has more than 15 years of experience in computational and experimental functional materials' physics. The particular emphasis is on the development of various technologies, including energy, water, and health. He is the first one in the scientific community to show that iron vanadate (FeVO4) is an intrinsic multiferroic system. Further, he has co-authored two books (i) Nanotechnology for Defence Applications (Springer) and (ii) Nanotechnology for Rural Development (Elsevier) and co-edited two books on Solar Energy (Springer). He has more than a hundred articles of international repute and two patents on his credit. He is an Editorial Board member of the Elsevier journal "Vacuum" and a life member of various societies such as Materials Research Society of India, Magnetic Society of India, Indian Carbon Society, and an annual American member Physical Society since 2007 and IEEE since 2019.*